\input ./style/arxiv-general.cfg
\documentclass[MSNbibl,nameyear,dvips]{arxstspdf}
\makeatletter
   \@ifpackageloaded{graphicx}{}{\usepackage{graphicx}}
\makeatother
\usepackage{flushend}
\usepackage{stfloats}
\usepackage{url,breakurl}

\volume{30}
\issue{2}
\pubyear{2015}
\firstpage{170}
\lastpage{175}
\doi{10.1214/15-STS517}
\referstodoi{10.1214/14-STS487}
\docsubty{FLA}

\makeatletter
\newcommand{\iint}{\int\!\!\!\int}
\newcommand{\Sigmamat}{\bolds{\Sigma}}
\makeatother

\begin{document}
\begin{frontmatter}
\vspace*{12pt}
\title{Capturing Multivariate Spatial Dependence: Model, Estimate and
then Predict}
\runtitle{Comment}

\begin{aug}
\author[A]{\fnms{Noel} \snm{Cressie}\corref{}\ead[label=e1]{ncressie@uow.edu.au}},
\author[A]{\fnms{Sandy} \snm{Burden}\ead[label=e2]{sburden@uow.edu.au}},
\author[A]{\fnms{Walter} \snm{Davis}\ead[label=e3]{walterd@uow.edu.au}},
\author[A]{\fnms{Pavel N.} \snm{Krivitsky}\ead[label=e4]{pavel@uow.edu.au}},
\author[A]{\fnms{Payam} \snm{Mokhtarian}\ead[label=e5]{payam@uow.edu.au}},
\author[A]{\fnms{Thomas}~\snm{Suesse}\ead[label=e6]{tsuesse@uow.edu.au}}
\and
\author[A]{\fnms{Andrew} \snm{Zammit-Mangion}\ead[label=e7]{azm@uow.edu.au}}
\runauthor{N. Cressie et al.}
\affiliation{National Institute for Applied Statistics Research
Australia (NIASRA), School of Mathematics and Applied Statistics, University of Wollongong, Australia.}

\address[A]{Noel~Cressie is Distinguished Professor,
Sandy~Burden is Research Fellow,
Walter~Davis is Senior Research Fellow,
Pavel N. Krivitsky is Lecturer,
Payam~Mokhtarian is Research Fellow,
Thomas~Suesse is Lecturer,
Andrew~Zammit-Mangion is Statistical Computing Scientist, National Institute for Applied Statistics Research
Australia (NIASRA), School of Mathematics and Applied Statistics, University of Wollongong, NSW 2522, Australia. \printead{e1},
\printead*{e2},
\printead*{e3},
\printead*{e4},
\printead*{e5},
\printead*{e6},
\printead*{e7}.}
\end{aug}

%
\begin{abstract}
Physical processes rarely occur in isolation, rather they influence and
interact with one another. Thus, there is great benefit in modeling
potential dependence between both spatial locations and different
processes. It is the interaction between these two dependencies that is
the focus of Genton and Kleiber's paper under discussion. We see the
problem of ensuring that any multivariate spatial covariance matrix is
nonnegative definite as important, but we also see it as a means to an
end. That ``end'' is solving the scientific problem of predicting a
multivariate field.
\end{abstract}

%
\begin{keyword}
\kwd{Asymmetric cross-covariance function}
\kwd{conditional approach}
\kwd{factor process}
\kwd{latent process}
\kwd{nonstationarity}
\end{keyword}
\end{frontmatter}

\section{Introduction}\label{sec:intro}
 We would like to thank Marc Genton and William Kleiber
(hereafter, GK) for their informative review of cross-covariance
functions in multivariate spatial statistics and the editor for the
opportunity to contribute to the discussion.

Physical processes rarely occur in isolation, rather they influence and
interact with one another. Thus, there is great benefit in modeling
potential dependence between both spatial locations and different
processes. It is the interaction between these two dependencies that is
the focus of GK.

We see the problem of ensuring that the matrix given in GK-(2) is
nonnegative definite (n.n.d.) as important, but we also see it as a means
to an end. That ``end'' is solving the scientific\vadjust{\goodbreak} problem of predicting
a multivariate field of, say, temperature and rainfall, based on noisy
and spatially incomplete data from weather stations in a region of
interest. There is also scientific interest in the behavior of the
measures of cross-spatial dependence (e.g., cross-covariance
functions), but usually spatial prediction is the ultimate goal.

Of course, an interim goal is estimation of the means, covariances and
cross-covariances, but not a lot of GK's review was devoted to this.
The nonparametric estimators given by GK-(6) and GK-(11) are useful for
recognizing which parametric \textit{class} of valid cross-covariance
functions might represent the multivariate spatial dependence in the
data. Estimation of the parameters in this class is usually obtained by
weighted least squares or maximum likelihood. Optimal spatial
prediction in practice proceeds by substituting these parameter
estimates into the model and computing the optimal data weights as if
the parameters were known. Because of this, predictors and their
standard errors are biased. These and other issues (e.g.,
change-of-support) are well known in the univariate spatial setting, and
they clearly also arise in the multivariate spatial setting.

One problem that arises in multivariate spatial statistics, but that is
not discussed very much by GK, is collocation (or not) of spatial data
from the different variables. This might be viewed as a missing-data
problem, for which a hierarchical multivariate spatial statistical
model offers a path forward.\vadjust{\goodbreak} Being hierarchical does not necessarily
mean being Bayesian, as the next section on latent modeling demonstrates.

\section{The Latent Process is Where the Spatial Dependence is Usually
Modeled}\label{sec:s1}
 The multivariate spatial models in GK do not account for the
measurement error that exists for all physical observations. Their
multivariate spatial processes are written as $\{\mathbf{ Z}(\mathbf{ s})\dvtx \mathbf{
s}\in\mathbb{R}^{d}\}$, and valid spatial-covariance models $\{\mathbf{
C}(\mathbf{ s}, \mathbf{ u})\dvtx \mathbf{ s},\mathbf{ u} \in\mathbb{R}^{d}\}$ are
constructed for them. GK's models are almost all ``smooth,'' in the
sense that
%
\begin{equation}
\label{eq:1} \lim_{\mathbf{ u}\to \mathbf{s}}\mathbf{ C}(\mathbf{ s, u})=\mathbf{
C}(\mathbf{ s, s}).
\end{equation}

However, an observation or potential observation is observed with
error, since no measuring instrument is perfect. Therefore, if the
observations are $\mathbf{ Z}(\mathbf{ s}_1), \ldots, \mathbf{ Z}(\mathbf{ s}_n)$,
then there is a hidden (or latent) process $\{\mathbf{ Y}(\mathbf{ s})\dvtx \mathbf{ s}
\in D\}$ such that
%
\begin{equation}
\label{eq:2} \mathbf{ Z}(\mathbf{ s}_{i})= \mathbf{ Y}(\mathbf{
s}_{i})+ \bolds{\varepsilon}(\mathbf{ s}_{i}),\quad i=1,\ldots,
n,
\end{equation}
where $\bolds{\varepsilon}(\mathbf{ s}_{i})$ has mean zero and
covariance matrix $\operatorname{cov}(\bolds{\varepsilon}(\mathbf{
s}_{i}))\equiv\Sigmamat_{\bolds\varepsilon}(\mathbf{ s}_{i})$.
Further, the measurement process is independent of the latent process,
and it is usually reasonable to assume that it is independent from one
observation to another, that is, $\operatorname{cov}(\bolds{\varepsilon}(\mathbf{ s}_{i}),\bolds{\varepsilon}(\mathbf{
s}_{j}))=\mathbf{0}$, for $i\neq j$.

It appears that GK build multivariate spatial covariance models for the
latent process, yet the definition of $\hat\mathbf{ C}(\mathbf{ h})$ in GK-(6)
is based on observations that always come with measurement error.
Hence, $\hat\mathbf{ C}(\mathbf{ h})$ is estimating a $\mathbf{ C}_{Z}(\mathbf{ h})$
that satisfies
%
\begin{equation}
\label{eq:3} \mathbf{ C}_{Z}(\mathbf{0})-\lim_{\mathbf{ h\to0}}
\mathbf{ C}_{Z}(\mathbf{ h}) \quad\mbox{is n.n.d.}
\end{equation}

The difference of the two matrices above is in fact the
measurement-error covariance matrix, which we denote as $\Sigmamat
_{\bolds\varepsilon}(\mathbf{0})$ in the stationary case
[i.e., where $\mathbf{ C}(\mathbf{ s},\mathbf{ s}+\mathbf{ h})$ depends only on $\mathbf{
h}$ and $\Sigmamat_{\bolds\varepsilon}(\mathbf{ s}) = \Sigmamat
_{\bolds\varepsilon}(\mathbf{0})$ for all $\mathbf{ s}$].

This mismatch between (\ref{eq:3}) and the stationary version of (\ref
{eq:1}), namely, $\lim_{\mathbf{ h\to0}}\mathbf{ C}(\mathbf{ h})=\mathbf{ C}(\mathbf{
0})$, can be resolved once one realizes that GK are really building
models for a latent $\mathbf{ Y}$-process, and that a full multivariate
spatial covariance function for the $\mathbf{ Z}$-process (i.e., the
observations) is obtained by additionally modeling the
measurement-error covariance, $\Sigmamat_{\bolds\varepsilon
}(\mathbf{
0})$. In their second example (GK-Section~6.2), GK recognize the need
for $\Sigmamat_{\bolds\varepsilon}(\mathbf{0})$: ``Due to the fact
that the data are observational, we augment each process' covariance
with a nugget effect.'' However, there are potentially nonzero
off-diagonal terms in $\Sigmamat_{\bolds\varepsilon}(\mathbf{0})$.
Notice that the two variables in GK-Section~6.2 are observed maximum
and minimum temperatures obtained from the \textit{same} instrument at
location $\mathbf{ s}_i$, say, and, hence, the measurement error for the
maximum, $\varepsilon_1(\mathbf{ s}_{i})$, and the measurement error for
the minimum, $\varepsilon_2(\mathbf{ s}_{i})$, should be correlated. GK's
choice of a diagonal matrix for $\Sigmamat_{\bolds\varepsilon
}(\mathbf{0})$ does not reflect this.

While it may not be obvious, there are also circumstances where a
``measurement error'' component is needed when modeling deterministic
spatio-temporal output from computer experiments, such as those used in
GK-Section~6.1. This component is actually a spatio-temporal
interaction that ``hides'' the latent spatial process (\cite{kang13}).

Finally, it is possible that a latent $\mathbf{ Y}$-process is itself not
``smooth.'' In this case, we can write the latent process as
%
\begin{equation}
\label{eq:5} \mathbf{ Y}(\mathbf{ s})=\mathbf{ W}(\mathbf{ s})+\bolds{\xi}(
\mathbf{ s}),\quad \mathbf{ s}\in D,
\end{equation}
where the $\mathbf{ W}$-process is smooth [i.e., satisfies (\ref{eq:1})]
and where the ${\bolds\xi}$-process is independent of the $\mathbf{
W}$-process and has mean zero. Often, it is assumed that ${\bolds
\xi}( \mathbf{ s})$ is independent of ${\bolds\xi}(\mathbf{ u})$ for any
$\mathbf{ s}\neq \mathbf{ u}$ and, when $\mathbf{ u}=\mathbf{ s},  \operatorname{cov}
(\bolds{\xi}(\mathbf{ s}),\bolds{\xi}(\mathbf{
u}))\equiv
\Sigmamat_{\bolds\xi}(\mathbf{0})$.

When modeling univariate spatial data, there has been considerable
inconsistency in the literature regarding how to handle the $\xi
$-process (micro-scale variation) and the $\varepsilon$-process
(measurement-error process). That confusion should also be avoided in
the multivariate spatial setting. We need both $\Sigmamat_{\bolds
\xi}(\mathbf{0})$ and $\Sigmamat_{\bolds\varepsilon}(\mathbf{0})$ for
different purposes, and these different roles should be accounted for:
We wish to filter out the ${\bolds\varepsilon}$-process (since it
is extraneous to the true, hidden $\mathbf{ Y}$-process), but we wish to
predict the ${\bolds\xi}$-process (since it represents the
scientific process at micro-scales). The presence of both processes is
manifested in $\hat\mathbf{ C}(\mathbf{ h})$, for $\mathbf h$ near $\mathbf0$, but
without more information than that supplied by the multivariate spatial
data, the $\bolds\xi$-process and the $\bolds\varepsilon
$-process are confounded.

\section{Estimate Using Cross-Variograms, then Predict Using
Cross-Covariances}\label{sec:s2}
 A small amount of GK's review of cross-covariances discusses
cross-variograms and generalized covariance functions, since they are
stationary when the process $\{\mathbf{ Z}(\mathbf{ s}) \dvtx \mathbf{ s} \in D\}$ is
differenced. Before differencing, the process is nonstationary. There
are a number of ways to do the differencing in a multivariate context,
leading to a lack of agreement among researchers of how to capture the
cross-dependence between the processes $\{Z_{q}(\mathbf{ s}) \dvtx \mathbf{ s} \in
D\}$ and $\{Z_{r}(\mathbf{ u})\dvtx \mathbf{ u}\in D\}, 1\le q\neq r\le p$. For
many scientific purposes, the key goal is optimal multivariate spatial
prediction. Therefore, the key measure of multivariate spatial
dependence should be one that can be used, without fail, in kriging and
co-kriging (i.e., spatial prediction) equations. That is, the optimal weights in the linear
combination of the spatial data, $\{Z_q(\mathbf{ s}_{qi})\dvtx i=1,\ldots
,n_{q}, q=1, \ldots, p\}$, should depend on this measure. If a measure
sometimes yields nonoptimal weights, we suggest that it is not as
interesting as one that does. While GK-(1) and GK-(4) yield optimal
weights, the (covariance-based) cross-variograms given by GK-(3) do not always.

\citeauthor{hoef93} (\citeyear{hoef93,hoef94}) give an example where
use of GK-(3) in spatial
multivariate prediction yields nonoptimal weights. Indeed, it is GK-(3)
that should have been tagged ``pseudo'' in the literature, not GK-(4).
The article that gives the most general multivariate spatial dependence
measure that is a function of $\mathbf{ h}=\mathbf{ s}-\mathbf{ u}$ is \citet
{kuns97}, who define the generalized cross-covariance functions.
Certainly, researchers' familiarity with these more general forms of
stationary cross-dependence is not high, but the interpretation of the
appropriate cross-variograms given by GK-(4) is not difficult (\cite{kuns97,maju97,cres98,huan09}).

We have found that for \textit{univariate} spatial processes, the
estimation of spatial-dependence parameters is achieved more stably
through the variogram than the covariance function (\cite{cres93}, Section~2.4.1). On the other hand, because optimal spatial prediction
using a valid covariance function can be used without fail (\cite{cres11}, Section~4.1.2), we recommend the following inferential
strategy for \textit{multivariate} spatial processes: Put the
cross-variogram at the core of parameter estimation and the
cross-covariance function at the core of optimal multivariate spatial
prediction.

We conclude that, for multivariate spatial processes, the appropriate
cross-variograms given by GK-(4) have great potential for the purpose
of estimation, and there is a need for a research program to pursue the
interpretation and robust estimation of GK-(4), but \textit{not} of the
inappropriate GK-(3).

\section{Capturing Spatial Dependence: A~Conditional Approach}\label{sec:s3}
 Let $[\cdot]$ denote the probability distribution of the
argument within square brackets. GK tackle cross-covariance
construction from the perspective of the \emph{joint} distribution,
$[Z_1(\cdot),Z_2(\cdot)]$. The \emph{conditional} approach to
constructing cross-covariance functions writes the joint distribution
as the product, $[Z_2(\cdot)| Z_1(\cdot)][Z_1(\cdot)]$. Consider the
space $D$ discretized onto a fine-resolution grid, $\{\mathbf{
s}_{1},\ldots
,\mathbf{ s}_{n}\}$, such that the processes $Z_{1}(\cdot)$ and
$Z_{2}(\cdot
)$ are represented as $n$-dimensional vectors $\mathbf{ Z}_{1}$ and $\mathbf{
Z}_{2}$, respectively. In practice, this is how a continuously indexed
process is represented in a computer program. Then, from \citeauthor{cres11} (\citeyear{cres11}, page
160), the conditional approach yields the bivariate spatial model,
%
\begin{eqnarray}
\operatorname{cov}(\mathbf{ Z}_{2})&=&\Sigmamat_{2|1}+\mathbf{
B}\Sigmamat_{11}\mathbf{ B}^{\prime},\label{eq:6-1}
\\
\operatorname{cov}(\mathbf{ Z}_{1},\mathbf{ Z}_{2})&=&
\Sigmamat_{11}\mathbf{ B}^{\prime
},\label{eq:6-2}
\\
\operatorname{cov}(\mathbf{ Z}_{1})&=&\Sigmamat_{11},\label{eq:6-3}
\end{eqnarray}
where $\Sigmamat_{2|1}$ and $\Sigmamat_{11}$ are n.n.d. matrices obtained
from univariate spatial processes, and the $n\times n$ matrix $\mathbf{ B}$
of real-valued entries is unrestricted. Critically, the joint matrix,
$\operatorname{cov} ((\mathbf{ Z}_{1}^{\prime},\mathbf{ Z}_{2}^{\prime
})^{\prime
} )$, is always n.n.d.

 Equations (\ref{eq:6-1})--(\ref{eq:6-3}) are obtained from
%
\begin{eqnarray}
\mathrm{E}(\mathbf{ Z}_{2}|\mathbf{ Z}_{1})&=&\mathbf{ B}
\mathbf{ Z}_{1},\label{eq:7}
\\
\operatorname{cov}(\mathbf{ Z}_{2}|\mathbf{ Z}_{1})&=&
\Sigmamat_{2|1},\label{eq:8}
\\
\operatorname{cov}(\mathbf{ Z}_{1})&=&\Sigmamat_{11},\label{eq:8-2}
\end{eqnarray}
which only involve univariate spatial processes. It should be
emphasized that the conditioning in (\ref{eq:7}) and (\ref{eq:8}) is on
the whole process $\mathbf{ Z}_{1}$. In contrast to what has been stated
elsewhere (\cite{bane15}, page 273), there is no attempt in the
conditional approach to build a joint distribution solely from
$Z_{2}(\mathbf{ s}_{i})|Z_{1}(\mathbf{ s}_{i})$, for $i=1,\ldots,n$. Indeed,
\[
\bigl[Z_{2}(\mathbf{ s}_{i})|Z_{1}(\mathbf{
s}_{i}) \bigr]=\int\cdots\int \frac
{ [Z_{2}(\mathbf{ s}_{i})|\mathbf{ Z}_{1}  ][\mathbf{
Z}_{1}]}{[Z_{1}(\mathbf{
s}_{i})]}\,d\mathbf{
Z}_{1,-i},
\]
where
\begin{eqnarray*}
&&d\mathbf{ Z}_{1,-i}\\
&&\quad=dZ_{1}(\mathbf{ s}_{1})\cdots
\,dZ_{1}(\mathbf{ s}_{i-1})\,dZ_{1}(\mathbf{
s}_{i+1})\cdots \,dZ_{1}(\mathbf{ s}_{n}).
\end{eqnarray*}

The order of the variables $\mathbf{ Z}_{1}$ and $\mathbf{ Z}_{2}$ in the
conditional approach is a choice that is generally driven by the
underlying science (e.g., \cite{royl99}). When more variables are
involved, the order may not always be obvious, but, if the goal is to
construct valid covariance and cross-covariance functions, the
different orderings can be viewed as enlarging the space of valid models.

\section{Capturing Spatial Dependence Through ``Factor'' Processes: A
Joint Approach}\label{sec:s4}
 Any univariate covariance function, $C(\mathbf{ s},\mathbf{ u})$,
that satisfies mild integrability conditions has a Karhunen--Lo\'{e}ve
representation (\cite{papo91}):
\[
C(\mathbf{ s},\mathbf{ u})=\sum_{a=1}^{\infty}
\lambda_{a}P_{a}(\mathbf{ s})P_{a}(\mathbf{ u}),
\]
where $\lambda_{1}\geq\lambda_{2}\geq\cdots\geq0$ and $ \{
P_{a}(\cdot)\dvtx a=1,2,\ldots \}$ are orthogonal eigenvectors
obtained by solving a Fredholm integral equation. After truncation, the
function,
%
\begin{equation}
\label{eq:3-1} C^{(b)}(\mathbf{ s},\mathbf{ u})\equiv\sum
_{a=1}^{b}\lambda_{a}P_{a}(
\mathbf{ s})P_{a}(\mathbf{ u}),
\end{equation}
is still n.n.d. Indeed, an equivalent way to write (\ref{eq:3-1}) is in
terms of a spatial process,
%
\begin{equation}
\label{eq:3-2} Z^{(b)}(\mathbf{ s})\equiv\mu(\mathbf{s})+\sum
_{a=1}^{b}\eta_{a}P_{a}(\mathbf{
s}),
\end{equation}
where ${\bolds\eta}\equiv(\eta_{1},\ldots,\eta_{b})^{\prime
}$ is
a mean-zero random vector with $\operatorname{cov}({\bolds\eta})=\operatorname
{diag}(\lambda_{1},\ldots,\lambda_{b})$.

Because the original covariance has been truncated, one way to capture
the lost covariation is to add back a simple random process:
%
\begin{equation}
\label{eq:3-3} Z(\mathbf{ s})\equiv\mu(\mathbf{s})+\sum_{a=1}^{b}
\eta_{a}P_{a}(\mathbf{ s})+\xi(\mathbf{ s}).
\end{equation}

It is common to choose $\xi(\cdot)$ to be a white-noise process, but it
is also straightforward to maintain some spatial structure in $\xi(\cdot)$
(\cite{berl00}). Clearly, the expression (\ref{eq:3-3}) could be thought
of as a spatial-factor-analysis model (\cite{chri01,lope08}), although
there are important differences in what is assumed known and what is estimated.

The definition (\ref{eq:3-3}) is directly expressed in terms of the
random components of the model, and it is a very fertile way of
constructing covariance functions: Specifically, replace $ \{
P_{a}(\cdot) \}$ with any set of known basis functions $ \{
S_{a}(\cdot) \}$, orthogonal or not; replace $\operatorname
{cov}({\bolds\eta})=\operatorname{diag}(\lambda_{1},\ldots,\lambda
_{b})$ with any $b\times b$ positive definite (p.d.) matrix $\mathbf K$; and
write $\operatorname{cov}(\xi(\mathbf{ s}),\xi(\mathbf{ u}))=\sigma_{\xi
}^{2}I(\mathbf{
u}=\mathbf{ s})$. Then
%
\begin{equation}
\label{eq:3-4} C(\mathbf{ s},\mathbf{ u})\equiv\mathbf{ S}(\mathbf{
s})^{\prime}\mathbf{ K}\mathbf{ S}(\mathbf{ u})+\sigma_{\xi}^{2}I(
\mathbf{ u}=\mathbf{ s})
\end{equation}
is a valid nonstationary univariate covariance model, where $\mathbf{
S}(\cdot)\equiv (S_{1}(\cdot),\ldots,S_{b}(\cdot)
)^{\prime}$.
\citet{cres08} call this a Spatial Random Effects (SRE) model.

The generalization of (\ref{eq:3-4}) to multivariate spatial processes
is easiest to obtain from its expression in terms of random components.
Here, the bivariate case shows its potential:
%
\begin{eqnarray}
\label{eq:3-5} %
Z_{1}(\mathbf{ s})&=&\mathbf{
S}^{(1)}(\mathbf{ s})^{\prime}{\bolds\eta }_{1}+
\xi_{1}(\mathbf{ s}),\quad \mathbf{ s}\in D,
\nonumber
\\[-8pt]
\\[-8pt]
\nonumber
Z_{2}(\mathbf{ s})&=&\mathbf{ S}^{(2)}(\mathbf{
s})^{\prime}{\bolds\eta }_{2}+\xi_{2}(\mathbf{ s}),\quad
\mathbf{ s}\in D,
\end{eqnarray}
where $\mathbf{ S}^{(1)}(\cdot)$ and $\mathbf{ S}^{(2)}(\cdot)$ are given
spatial basis functions that are quite likely to be different for
$Z_{1}(\cdot)$ and for $Z_{2}(\cdot)$, and ${\bolds\eta}_{1}$ and
${\bolds\eta}_{2}$ may have nonzero $\operatorname{cov}({\bolds
\eta}_{1},{\bolds\eta}_{2})$; see \citet{brad14}. Note that
(\ref
{eq:3-5}) could be viewed as an errors-in-variables parameterization
(Christensen\break and Amemiya, \citeyear{chri02,chri03}). Clearly, the implied covariance and
cross-covariance functions are nonstationary, but their parameters can
still be estimated from the multivariate spatial data.

Representations using ``factor'' processes, such as in~(\ref{eq:3-5}),
generalize many of the constructions outlined in GK-Section~2. Let
$ \{U_{r}(\cdot)\dvtx r=1,\ldots,p \}$ be a set of independent
univariate processes with mean $0$, variance $1$ and stationary
correlation functions $\{\rho_{r}(\mathbf{ h})\}$. Suppose further that
$ \{g_{qr}(\mathbf{ h})\dvtx q,r=1,\ldots,p \}$ are integrable
kernels; then a very general factor representation is
%
\begin{equation}
\label{eq:3-6} Z_{q}(\mathbf{ s})\equiv\sum
_{r=1}^{p}\int g_{qr}(\mathbf{ u}-\mathbf{
s})U_{r}(\mathbf{ u})\,d\mathbf{ u}.
\end{equation}

 The cross-covariances implied by (\ref{eq:3-6}) are
\begin{eqnarray*}
\label{eq:3-7} && C_{qr}(\mathbf{ h})\\
&&\quad \equiv\sum
_{k=1}^{p}\iint g_{qk}(\mathbf{
v}_{1})g_{rk}(\mathbf{ v}_{2})\rho_{k}(
\mathbf{ v}_{1}-\mathbf{ v}_{2}+\mathbf{ h})\,d\mathbf{
v}_{1}\,d\mathbf{ v}_{2}.
\end{eqnarray*}

Clearly, asymmetry is present in all but the simplest cases, and the
linear model of coregionalization in GK-Section~2.1 is recovered by
setting $g_{qr}(\mathbf{ u}-\mathbf{ s})=A_{qr}\delta(\mathbf{ u}-\mathbf{ s})$ in
(\ref
{eq:3-6}), where $\delta(\cdot)$ is the Dirac delta function. The
cross-covariance function given at the beginning of GK-Section~2.2 is
also recovered by setting $\rho_{k}(\mathbf{ h})\equiv\rho(\mathbf{ h})/p$,
$g_{qk}(\mathbf{ h})\equiv k_{q}(\mathbf{ h})$, and $g_{rk}(\mathbf{ h})\equiv
k_{r}(\mathbf{ h})$. There are many familiar special cases, and the
``factor'' processes $\{U_{r}(\cdot)\}$ in (\ref{eq:3-6}) do not even
need to be independent.

\section{Comments on the Data Examples}\label{sec:s5}
 In GK-Section~6, several bivariate spatial models are
implemented on data describing pressure and temperature. GK compare
these models and assess which ones are best able to capture the
dependence within and between the processes. This may be the first time
such an exercise has been carried out; nevertheless, there are aspects
of the analyses we would modify.

First, the various models under consideration contain different numbers
of free parameters. For example, the parsimonious Mat{\'e}rn has six
free parameters, while the nonstationary parsimonious Mat{\'e}rn has
several hundred. In this context, the log-likelihood does not provide a
useful comparison of model fit. We suggest that the Akaike Information
Criterion (AIC) and its corrected version (AICc) are preferable when
the number of free parameters differ across models (\cite{Hoeting2006,Lee2009}).

Second, the other summaries (RMSE and CRPS) may not be indicative of
model performance, since parameters estimated from the entire dataset
were used. In our view, the flexibility, adaptability\
and utility of a model can only be assessed (in the context of
cross-validation) using data that have not been used for parameter estimation.

Finally, as we mentioned earlier (Section~\ref{sec:s1}), the so-called
nugget-effect matrix that consists of both measurement-error and
micro-scale matrices needs to be modeled in both examples, and in the
second example (GK-Section~6.2) the possibility of nondiagonal
contributions should be considered.

\section{Bibliographic Notes}\label{sec:s6}
 It is a big task to review multivariate geostatistics, and we
appreciate that GK had limits on what they could cover. They selected a
few topics that went beyond their core goal of reviewing
cross-covariance functions and to some of these topics we add the
following bibliographic notes.
\subsection*{Nonstationarity for Factor Processes}
\citeauthor{wikl01} (\citeyear{wikl10}): Recall from Section~\ref{sec:s4} that, in the
univariate setting, reduced-rank covariance functions are very useful
for big spatial data:
\begin{eqnarray*}
\operatorname{cov}\bigl(Z(\mathbf{ s}),Z(\mathbf{ u})\bigr)=\mathbf{ S}(\mathbf{
s})^{\prime}\mathbf{ K}\mathbf{ S}(\mathbf{ u})+v(\mathbf{ s})I(\mathbf{ u}=
\mathbf{ s}),
\end{eqnarray*}
where $\mathbf{ S}(\cdot)$ is a given $b$-dimensional $(b\ll n)$ vector of
spatial basis functions, $\mathbf{ K}$ is an unknown $b\times b$ p.d. matrix,
and $v(\mathbf{ s})> 0$. In Wikle's review of these rank-$b$ covariance
models, he points out their computational advantages for spatial
prediction. A generalization to multivariate covariance functions is
straightforward; see Section~\ref{sec:s4}. When considering global
\mbox{processes}, such as in remote sensing applications, the nonstationarity
of these models is an advantage.

\subsection*{Asymmetric Cross-Covariance Functions}
 The asymmetry in multivariate spatial processes may come
from, say, preferable mineralization of metals in an ore body or
deposition of lighter particulate matter in the environment.
Cross-covariance models should be able to detect such phenomena. The
``shifted-lag model'' is a natural way to capture sources of asymmetry
and has a longer history than that indicated by the reference to \citet
{lizh11} in GK-Section~5.1, as illustrated by the following articles.

\citeauthor{hoef93} (\citeyear{hoef93,hoef94}): The shifted-lag model given in GK-(12) was proposed.

\citet{maju97}: The shifted-lag model was estimated from variance-based
cross-variograms.

\citeauthor{chri01} (\citeyear{chri01,chri02}):
A latent variable factor analysis model for a
multivariate spatial process was based on the shifted-lag model.

\subsection*{Spatio-Temporal Covariance Functions}
From the point of view of building covariance functions, the
temporal dimension could be viewed simply as an extra ``spatial''
dimension. However, an alternative approach, based on the dynamical
evolution of spatial processes in time, often allows optimal prediction
to be carried out without explicitly constructing covariance models.

\citet{wikl01}: This article uses conditional-probability modeling (in
space and time) and reduced-rank models to achieve optimal
spatio-temporal prediction. It bypasses the need for constructing
spatio-temporal cross-covariance functions.

\citet{cres11}: In much of this book, cross-covariance functions are
considered as derivative measures, from scientifically interpretable
dynamical (multivariate) spatial models; see their pages 418--425. A
hierarchical dynamical approach is taken that yields optimal
spatio-temporal prediction directly, without having to pass through
multivariate covariance-function modeling.

In our view, hierarchical physical-statistical modeling of big,
spatio-temporal, multivariate, nonlinear, non-Gaussian data will
represent the next frontier.


\begin{thebibliography}{24}

\bibitem[\protect\citeauthoryear{Banerjee, Carlin and Gelfand}{2015}]{bane15}
%
\begin{bbook}[author]
\bauthor{\bsnm{Banerjee},~\bfnm{S.}\binits{S.}},
\bauthor{\bsnm{Carlin},~\bfnm{B.~P.}\binits{B.~P.}} \AND
\bauthor{\bsnm{Gelfand},~\bfnm{A.~E.}\binits{A.~E.}}
(\byear{2015}).
\btitle{Hierarchical Modeling and Analysis for Spatial Data},
\bedition{2nd} ed.
\bpublisher{Chapman \& Hall/CRC},
\blocation{Boca Raton, FL}.
\end{bbook}
%

\bptok{imsref}%
\endbibitem

\bibitem[\protect\citeauthoryear{Berliner, Wikle and Cressie}{2000}]{berl00}
%
\begin{barticle}[author]
\bauthor{\bsnm{Berliner},~\bfnm{L.~M.}\binits{L.~M.}},
\bauthor{\bsnm{Wikle},~\bfnm{C.~K.}\binits{C.~K.}} \AND
\bauthor{\bsnm{Cressie},~\bfnm{N.}\binits{N.}}
(\byear{2000}).
\btitle{Long-lead prediction of pacific SSTs via Bayesian dynamic modeling}.
\bjournal{Journal of Climate}
\bvolume{13}
\bpages{3953--3968}.
\end{barticle}
%

\bptok{imsref}%
\endbibitem

\bibitem[\protect\citeauthoryear{Bradley, Holan and Wikle}{2015}]{brad14}
%
\begin{bmisc}[author]
\bauthor{\bsnm{Bradley},~\bfnm{J.~R.}\binits{J.~R.}},
\bauthor{\bsnm{Holan},~\bfnm{S.~H.}\binits{S.~H.}} \AND
\bauthor{\bsnm{Wikle},~\bfnm{C.~K.}\binits{C.~K.}}
(\byear{2015}).
\bhowpublished{Multivariate spatio-temporal models for high-dimensional
areal data with application to longitudinal employer-household dynamics
Available at \arxivurl{arXiv:1503.00982}.}
\end{bmisc}
%

\bptok{imsref}%
\endbibitem

\bibitem[\protect\citeauthoryear{Christensen and Amemiya}{2001}]{chri01}
%
\begin{barticle}[mr]
\bauthor{\bsnm{Christensen},~\bfnm{William~F.}\binits{W.~F.}} \AND
\bauthor{\bsnm{Amemiya},~\bfnm{Yasuo}\binits{Y.}}
(\byear{2001}).
\btitle{Generalized shifted-factor analysis method for multivariate
geo-referenced data}.
\bjournal{Math. Geol.}
\bvolume{33}
\bpages{801--824}.
\bid{doi={10.1023/A:1010998730645}, issn={0882-8121}, mr={1976053}}
\end{barticle}
%

\bptok{imsref}%
\endbibitem

\bibitem[\protect\citeauthoryear{Christensen and Amemiya}{2002}]{chri02}
%
\begin{barticle}[mr]
\bauthor{\bsnm{Christensen},~\bfnm{William~F.}\binits{W.~F.}} \AND
\bauthor{\bsnm{Amemiya},~\bfnm{Yasuo}\binits{Y.}}
(\byear{2002}).
\btitle{Latent variable analysis of multivariate spatial data}.
\bjournal{J. Amer. Statist. Assoc.}
\bvolume{97}
\bpages{302--317}.
\bid{doi={10.1198/016214502753479437}, issn={0162-1459}, mr={1947288}}
\end{barticle}
%

\bptok{imsref}%
\endbibitem

\bibitem[\protect\citeauthoryear{Christensen and Amemiya}{2003}]{chri03}
%
\begin{barticle}[mr]
\bauthor{\bsnm{Christensen},~\bfnm{William~F.}\binits{W.~F.}} \AND
\bauthor{\bsnm{Amemiya},~\bfnm{Yasuo}\binits{Y.}}
(\byear{2003}).
\btitle{Modeling and prediction for multivariate spatial factor analysis}.
\bjournal{J. Statist. Plann. Inference}
\bvolume{115}
\bpages{543--564}.
\bid{doi={10.1016/S0378-3758(02)00173-8}, issn={0378-3758}, mr={1985883}}
\end{barticle}
%

\bptok{imsref}%
\endbibitem

\bibitem[\protect\citeauthoryear{Cressie}{1993}]{cres93}
%
\begin{bbook}[mr]
\bauthor{\bsnm{Cressie},~\bfnm{Noel~A.~C.}\binits{N.~A.~C.}}
(\byear{1993}).
\btitle{Statistics for Spatial Data},
\bedition{rev}~ed.
\bpublisher{Wiley},
\blocation{New York}.
\bid{mr={1239641}}
\end{bbook}
%

\bptok{imsref}%
\endbibitem

\bibitem[\protect\citeauthoryear{Cressie and Johannesson}{2008}]{cres08}
%
\begin{barticle}[mr]
\bauthor{\bsnm{Cressie},~\bfnm{Noel}\binits{N.}} \AND
\bauthor{\bsnm{Johannesson},~\bfnm{Gardar}\binits{G.}}
(\byear{2008}).
\btitle{Fixed rank kriging for very large spatial data sets}.
\bjournal{J. R. Stat. Soc. Ser. B. Stat. Methodol.}
\bvolume{70}
\bpages{209--226}.
\bid{doi={10.1111/j.1467-9868.2007.00633.x}, issn={1369-7412}, mr={2412639}}
\end{barticle}
%

\bptok{imsref}%
\endbibitem

\bibitem[\protect\citeauthoryear{Cressie and Wikle}{1998}]{cres98}
%
\begin{barticle}[mr]
\bauthor{\bsnm{Cressie},~\bfnm{Noel}\binits{N.}} \AND
\bauthor{\bsnm{Wikle},~\bfnm{Christopher~K.}\binits{C.~K.}}
(\byear{1998}).
\btitle{The variance-based cross-variogram: You can add apples and oranges}.
\bjournal{Math. Geol.}
\bvolume{30}
\bpages{789--799}.
\bid{doi={10.1023/A:1021770324434}, issn={0882-8121}, mr={1646242}}
\end{barticle}
%

\bptok{imsref}%
\endbibitem

\bibitem[\protect\citeauthoryear{Cressie and Wikle}{2011}]{cres11}
%
\begin{bbook}[mr]
\bauthor{\bsnm{Cressie},~\bfnm{Noel}\binits{N.}} \AND
\bauthor{\bsnm{Wikle},~\bfnm{Christopher~K.}\binits{C.~K.}}
(\byear{2011}).
\btitle{Statistics for Spatio-Temporal Data}.
\bpublisher{Wiley},
\blocation{Hoboken, NJ}.
\bid{mr={2848400}}
\end{bbook}
%

\bptok{imsref}%
\endbibitem

\bibitem[\protect\citeauthoryear{Hoeting et~al.}{2006}]{Hoeting2006}
%
\begin{barticle}[pbm]
\bauthor{\bsnm{Hoeting},~\bfnm{Jennifer~A.}\binits{J.~A.}},
\bauthor{\bsnm{Davis},~\bfnm{Richard~A.}\binits{R.~A.}},
\bauthor{\bsnm{Merton},~\bfnm{Andrew~A.}\binits{A.~A.}} \AND
\bauthor{\bsnm{Thompson},~\bfnm{Sandra~E.}\binits{S.~E.}}
(\byear{2006}).
\btitle{Model selection for geostatistical models}.
\bjournal{Ecol. Appl.}
\bvolume{16}
\bpages{87--98}.
\bid{issn={1051-0761}, pmid={16705963}}
\end{barticle}
%

\bptok{imsref}%
\endbibitem

\bibitem[\protect\citeauthoryear{Huang et~al.}{2009}]{huan09}
%
\begin{barticle}[mr]
\bauthor{\bsnm{Huang},~\bfnm{Chunfeng}\binits{C.}},
\bauthor{\bsnm{Yao},~\bfnm{Yonggang}\binits{Y.}},
\bauthor{\bsnm{Cressie},~\bfnm{Noel}\binits{N.}} \AND
\bauthor{\bsnm{Hsing},~\bfnm{Tailen}\binits{T.}}
(\byear{2009}).
\btitle{Multivariate intrinsic random functions for cokriging}.
\bjournal{Math. Geosci.}
\bvolume{41}
\bpages{887--904}.
\bid{doi={10.1007/s11004-009-9218-4}, issn={1874-8961}, mr={2557237}}
\end{barticle}
%

\bptok{imsref}%
\endbibitem

\bibitem[\protect\citeauthoryear{Kang and Cressie}{2013}]{kang13}
%
\begin{barticle}[author]
\bauthor{\bsnm{Kang},~\bfnm{E.~L.}\binits{E.~L.}} \AND
\bauthor{\bsnm{Cressie},~\bfnm{N.}\binits{N.}}
(\byear{2013}).
\btitle{Bayesian hierarchical ANOVA of regional climate-change
projections from NARCCAP phase II}.
\bjournal{International Journal of Applied Earth Observation and
Geoinformation}
\bvolume{22}
\bpages{3--15}.
\end{barticle}
%

\bptok{imsref}%
\endbibitem

\bibitem[\protect\citeauthoryear{K{\"u}nsch, Papritz and
Bassi}{1997}]{kuns97}
%
\begin{barticle}[mr]
\bauthor{\bsnm{K{\"u}nsch},~\bfnm{H.~R.}\binits{H.~R.}},
\bauthor{\bsnm{Papritz},~\bfnm{A.}\binits{A.}} \AND
\bauthor{\bsnm{Bassi},~\bfnm{F.}\binits{F.}}
(\byear{1997}).
\btitle{Generalized cross-covariances and their estimation}.
\bjournal{Math. Geol.}
\bvolume{29}
\bpages{779--799}.
\bid{doi={10.1007/BF02768902}, issn={0882-8121}, mr={1470669}}
\end{barticle}
%

\bptok{imsref}%
\endbibitem

\bibitem[\protect\citeauthoryear{Lee and Ghosh}{2009}]{Lee2009}
%
\begin{barticle}[mr]
\bauthor{\bsnm{Lee},~\bfnm{Hyeyoung}\binits{H.}} \AND
\bauthor{\bsnm{Ghosh},~\bfnm{Sujit~K.}\binits{S.~K.}}
(\byear{2009}).
\btitle{Performance of information criteria for spatial models}.
\bjournal{J. Stat. Comput. Simul.}
\bvolume{79}
\bpages{95--106}.
\bid{doi={10.1080/00949650701611143}, issn={0094-9655}, mr={2655676}}
\bptnote{check pages}%
\end{barticle}
%

\bptok{imsref}%
\endbibitem

\bibitem[\protect\citeauthoryear{Li and Zhang}{2011}]{lizh11}
%
\begin{barticle}[mr]
\bauthor{\bsnm{Li},~\bfnm{Bo}\binits{B.}} \AND
\bauthor{\bsnm{Zhang},~\bfnm{Hao}\binits{H.}}
(\byear{2011}).
\btitle{An approach to modeling asymmetric multivariate spatial
covariance structures}.
\bjournal{J. Multivariate Anal.}
\bvolume{102}
\bpages{1445--1453}.
\bid{doi={10.1016/j.jmva.2011.05.010}, issn={0047-259X}, mr={2819961}}
\end{barticle}
%

\bptok{imsref}%
\endbibitem

\bibitem[\protect\citeauthoryear{Lopes, Salazar and Gamerman}{2008}]{lope08}
%
\begin{barticle}[mr]
\bauthor{\bsnm{Lopes},~\bfnm{Hedibert~Freitas}\binits{H.~F.}},
\bauthor{\bsnm{Salazar},~\bfnm{Esther}\binits{E.}} \AND
\bauthor{\bsnm{Gamerman},~\bfnm{Dani}\binits{D.}}
(\byear{2008}).
\btitle{Spatial dynamic factor analysis}.
\bjournal{Bayesian Anal.}
\bvolume{3}
\bpages{759--792}.
\bid{doi={10.1214/08-BA329}, issn={1936-0975}, mr={2469799}}
\end{barticle}
%

\bptok{imsref}%
\endbibitem

\bibitem[\protect\citeauthoryear{Majure and Cressie}{1997}]{maju97}
%
\begin{barticle}[author]
\bauthor{\bsnm{Majure},~\bfnm{J.}\binits{J.}} \AND
\bauthor{\bsnm{Cressie},~\bfnm{N.}\binits{N.}}
(\byear{1997}).
\btitle{Dynamic graphics for exploring spatial dependence in
multivariate spatial data}.
\bjournal{Geographical Systems}
\bvolume{4}
\bpages{131--158}.
\end{barticle}
%

\bptok{imsref}%
\endbibitem

\bibitem[\protect\citeauthoryear{Papoulis}{1991}]{papo91}
%
\begin{bbook}[author]
\bauthor{\bsnm{Papoulis},~\bfnm{A.}\binits{A.}}
(\byear{1991}).
\btitle{Probability, Random Variables, and Stochastic Processes}.
\bpublisher{McGraw-Hill},
\blocation{Columbus, OH}.
\end{bbook}
%

\bptok{imsref}%
\endbibitem

\bibitem[\protect\citeauthoryear{Royle et~al.}{1999}]{royl99}
%
\begin{bincollection}[author]
\bauthor{\bsnm{Royle},~\bfnm{J.~A.}\binits{J.~A.}},
\bauthor{\bsnm{Berliner},~\bfnm{L.~M.}\binits{L.~M.}},
\bauthor{\bsnm{Wikle},~\bfnm{C.~K.}\binits{C.~K.}} \AND
\bauthor{\bsnm{Milliff},~\bfnm{R.}\binits{R.}}
(\byear{1999}).
\btitle{A hierarchical spatial model for constructing wind fields from
scatterometer data in the Labrador Sea}.
In \bbooktitle{Case Studies in Bayesian Statistics IV}
(\beditor{\bfnm{R.~E.}\binits{R.~E.}~\bsnm{Gatsonis}},
\beditor{\bfnm{B.}\binits{B.}~\bsnm{Kass}},
\beditor{\bfnm{A.}\binits{A.}~\bsnm{Carlin}},
\beditor{\bfnm{A.}\binits{A.}~\bsnm{Carriquiry}},
\beditor{\bfnm{I.}\binits{I.}~\bsnm{Gelman}} \AND
\beditor{\bfnm{M.}\binits{M.}~\bsnm{Verdinelli}}, eds.)
\bpages{367--382}.
\bpublisher{Springer},
\blocation{New York, NY}.
\end{bincollection}
%

\bptok{imsref}%
\endbibitem

\bibitem[\protect\citeauthoryear{Ver~Hoef and Cressie}{1993}]{hoef93}
%
\begin{barticle}[mr]
\bauthor{\bsnm{Ver Hoef},~\bfnm{Jay~M.}\binits{J.~M.}} \AND
\bauthor{\bsnm{Cressie},~\bfnm{Noel}\binits{N.}}
(\byear{1993}).
\btitle{Multivariable spatial prediction}.
\bjournal{Math. Geol.}
\bvolume{25}
\bpages{219--240}.
\bid{doi={10.1007/BF00893273}, issn={0882-8121}, mr={1206187}}
\end{barticle}
%

\bptok{imsref}%
\endbibitem

\bibitem[\protect\citeauthoryear{Ver~Hoef and Cressie}{1994}]{hoef94}
%
\begin{barticle}[author]
\bauthor{\bsnm{Ver~Hoef},~\bfnm{J.~M.}\binits{J.~M.}} \AND
\bauthor{\bsnm{Cressie},~\bfnm{N.}\binits{N.}}
(\byear{1994}).
\btitle{Errata: Multivariable spatial prediction}.
\bjournal{Math. Geol.}
\bvolume{26}
\bpages{273--275}.
\end{barticle}
%

\bptok{imsref}%
\endbibitem

\bibitem[\protect\citeauthoryear{Wikle}{2010}]{wikl10}
%
\begin{bincollection}[mr]
\bauthor{\bsnm{Wikle},~\bfnm{Christopher~K.}\binits{C.~K.}}
(\byear{2010}).
\btitle{Low-rank representations for spatial processes}.
In \bbooktitle{Handbook of Spatial Statistics}
\bpages{107--118}.
\bpublisher{CRC Press},
\blocation{Boca Raton, FL}.
\bid{doi={10.1201/9781420072884-c8}, mr={2730946}}
\end{bincollection}
%

\bptok{imsref}%
\endbibitem

\bibitem[\protect\citeauthoryear{Wikle et~al.}{2001}]{wikl01}
%
\begin{barticle}[mr]
\bauthor{\bsnm{Wikle},~\bfnm{Christopher~K.}\binits{C.~K.}},
\bauthor{\bsnm{Milliff},~\bfnm{Ralph~F.}\binits{R.~F.}},
\bauthor{\bsnm{Nychka},~\bfnm{Doug}\binits{D.}} \AND
\bauthor{\bsnm{Berliner},~\bfnm{L.~Mark}\binits{L.~M.}}
(\byear{2001}).
\btitle{Spatiotemporal hierarchical {B}ayesian modeling: Tropical
ocean surface winds}.
\bjournal{J. Amer. Statist. Assoc.}
\bvolume{96}
\bpages{382--397}.
\bid{doi={10.1198/016214501753168109}, issn={0162-1459}, mr={1939342}}
\end{barticle}
%
\bptok{imsref}%
\endbibitem
\end{thebibliography}

%

%



\end{document}